# Gravitational Waves I: Basic Emission Equations.

M. Cattani
Institute of Physics, University of S. Paulo, C.P. 66318, CEP 05315-970
S. Paulo, S.P. Brazil .
E−mail: mcattani@if.usp.br

Abstract.
　　Using the Einstein gravitation theory we show how to obtain the basic equations which predict the gravitational waves. This paper was written to graduate and post-graduate students of Physics. We deduce the equations didactically following the simplest way maintaining, however, the necessary mathematical rigor.

Resumo.
　　Usando a teoria de gravitação de Einstein mostramos como obter as equações básicas que prevêem as ondas gravitacionais. Este artigo foi escrito visando estudantes de graduação e de pós−graduação em Física. As equações são deduzidas didaticamente seguindo um caminho mais simples possível mantendo, entretanto, o necessário rigor matemático.

## 1. Introdução. As Equações de Campo de Einstein.

　　Em um artigo anterior publicado na Revista Brasileira de Ensino de Física[1] mostramos como deduzir de modo didático e mais simples possível, sem perder o rigor matemático, as Equações da Teoria de Teoria de Gravitação de Einstein. Esse artigo foi destinado aos alunos de graduação e de pós−graduação de Física.

　　Com o mesmo intuito, escrevemos o presente artigo (indicado por I) para mostrar como obter as *equações básicas* que prevêem as Ondas Gravitacionais partindo das Equações de Campo de Einstein. Como muitos trabalhos e livros excelentes foram escritos sobre a Teoria de Gravitação de Einstein e suas previsões sobre as ondas gravitacionais, não faremos uma revisão sobre o referido assunto. A nossa intenção é, simplesmente, citar e seguir somente uns poucos livros textos e artigos. Num próximo artigo (II) veremos como estimar as intensidades das ondas no caso de vários sistemas físicos[3-6](sistemas estelares binários, colapso gravitacional de estrelas, pulsares, etc...) e como detectá−las.

　　As Equações de Campo de Einstein, num espaço 4 − dim de Riemann são dadas por [1-6]

$$R_{\mu\nu} - (1/2)g_{\mu\nu} R = \kappa\, T_{\mu\nu}^{(m)} \qquad (1.1),$$

onde $\kappa = 8\pi G/c^4$, $R_{\mu\nu}$ é o *tensor de curvatura* de Ricci definido por

$$R_{\mu\nu} = R_{\nu\mu} = \partial_\mu \Gamma_{\nu\sigma}{}^\sigma - \partial_\sigma \Gamma_{\nu\mu}{}^\sigma + \Gamma_{\mu\sigma}{}^\tau \Gamma_{\nu\mu}{}^\sigma - \Gamma_{\nu\mu}{}^\tau \Gamma_{\tau\sigma}{}^\sigma \qquad (1.2),$$

$\Gamma$ são os *símbolos de Christoffel* dados por

$$\Gamma_{\mu\nu}{}^\alpha = (g^{\alpha\lambda}/2)(\partial_\nu g_{\lambda\mu} + \partial_\mu g_{\lambda\nu} - \partial_\lambda g_{\mu\nu}) \qquad (1.3),$$

$g_{\mu\nu}$ o *tensor métrico* do espaço onde o *elemento de linha* ds é definido por $ds^2 = g_{\mu\nu}(x)\, dx^\mu dx^\nu$ (*forma quadrática fundamental*) e $x^\lambda$ as coordenadas do espaço métrico. No caso de espaços curvos o tensor $g_{\mu\nu}$ é função das coordenadas $x^\lambda$. O escalar $R = g^{\lambda\sigma} R_{\sigma\lambda}$ é conhecido como *curvatura escalar* ou *invariante de curvatura* do espaço. Lembramos que nas equações acima os índices $\mu,\nu,... = 1,2,3,4$ e que seguimos a convenção de Einstein quando temos índices repetidos nas expressões. Na ausência da gravitação, o espaço (a arena) onde observamos os fenômenos físicos é o *espaço plano* de Minkowski (com curvatura nula, $R = 0$) onde $x^\lambda = (\mathbf{x}, x^4) = (\mathbf{x}, ct)$, $ds^2 = -(d\mathbf{x})^2 + (dx^4)^2 = c^2 dt^2 - d\mathbf{x}^2$ com tensor métrico $g_{\mu\nu}^{(o)} = (-1,-1,-1, 1)$. Finalmente, $T_{\mu\nu}^{(m)}$ é o tensor energia–momento da *matéria* que, no caso de um fluido ideal, é dado por [1-7]

$$T_{\mu\nu}^{(m)} = (\varepsilon + p)u_\mu u_\nu - g_{\mu\nu}\, p \qquad (1.4),$$

onde $\varepsilon$ é a densidade escalar *própria* de energia da matéria, p a pressão e $u_\alpha$ é o quadrivetor "*velocidade*" (*4–velocidade*) definido por $u_\alpha = dx_\alpha/ds$. Notemos que apesar de $T^\mu{}_\nu{}^{(m)}$ obedecer à equação $T^\mu{}_\nu{}^{(m)}{}_{;\mu} = 0$ ela não exprime nenhuma lei de conservação [ref.2, pág.307]. Quando a densidade de energia de repouso das partículas for preponderante, teremos $\varepsilon \approx \rho_o c^2$ onde $\rho_o$ é a soma das massas das partículas por unidade de volume (ou densidade de massa) medido no referencial em que as partículas estão em repouso, ou seja, é a densidade de massa. Note que s é um parâmetro puramente matemático. No caso particular de um espaço de Minkowski (Apêndice I) ele está associado ao *tempo próprio* $\tau$ [2-5] e a *4–velocidade* $u_\alpha$ está associada ao vetor velocidade $\mathbf{v} = d\mathbf{x}/dt$ em 3–dim (Apêndice II).

O 4–vetor "*aceleração*" $a^\alpha$ que é definido por $a^\alpha = du^\alpha/ds = d^2x^\alpha/ds^2$ obedece às equações de movimento denominadas de *equações de uma geodésica* [1-7]

$$d^2x^\alpha/ds^2 + \Gamma_{\tau\nu}{}^\alpha (dx^\nu/ds)(dx^\tau/ds) = 0 \qquad (1.5).$$

Através delas vemos que o movimento de uma partícula em um campo de gravitação é definido pelas grandezas $\Gamma_{\tau\nu}^{\alpha}$. Desse modo, podemos chamar a grandeza $-m\,\Gamma_{\tau\nu}^{\alpha}\,(dx^{\nu}/ds)\,(dx^{\tau}/ds)$ de "4–*força*" que age sobre a partícula num campo gravitacional. Assim, o tensor $g_{\mu\nu}$ desempenha o papel de "*potenciais*" do campo de gravitação e suas derivadas definem o "*campo de gravitação*" $\Gamma_{\mu\nu}^{\alpha}$.

Notemos que as equações de campo (1) constituem um sistema de dez equações de derivadas parciais, não lineares, de segunda ordem. Elas são constituídas por dez grandezas desconhecidas: seis componentes de $g_{\mu\nu}$, três componentes de $u_{\mu}$ e a densidade ρ da matéria (ou a pressão p). Lembrando que temos de conhecer de antemão a equação de estado da matéria que relaciona ρ e p.

Como as equações de campo da gravitação não são lineares, o princípio de superposição não é válido para os campos gravitacionais (a não ser para campos fracos como veremos na Secção 2), contrariamente ao que ocorre para os campos eletromagnéticos na relatividade restrita.

No caso do eletromagnetismo, uma dada distribuição de cargas determina, através das equações de Maxwell, o campo que elas criam. Esse campo, por sua vez, determinará o movimento das cargas. No caso da gravitação, isto não pode ser feito, pois o campo gravitacional e movimento da matéria são determinados concomitantemente por intermédio das equações (1), dadas as condições iniciais.

No desenvolvimento do artigo, que é composto por quatro Seções e três Apêndices adotamos o seguinte procedimento: colocamos nas Seções os resultados mais importantes sem nos preocuparmos muito com detalhes de cálculos. Estes são mostrados nos Apêndices.

Na Seção 2, no limite de campos gravitacionais fracos, linearizando as equações de campo (1.1), deduziremos a *equação fundamental* a partir da qual pode-se calcular a emissão de ondas gravitacionais. Na Seção 3, estudamos a propagação das ondas gravitacionais no vácuo e calculamos *fluxo de energia* que elas transportam e as *forças de cisalhamento* que as mesmas causam sobre corpos massivos. Finalmente, na Seção 4, usando a equação fundamental deduzida na Seção 2, calculamos as propriedades das ondas gravitacionais que são geradas por *quadrupolos de massa*.

## 2. Linearização das Equações de Einstein e as Ondas Gravitacionais.

Veremos agora como linearizar as (1.1) assumindo que o campo gravitacional gerado pelas massas seja fraco de tal modo que a métrica $g_{\mu\nu}$ do espaço–tempo difira muito pouco da métrica de Minkowski. Ou seja, $g_{\mu\nu} = g_{\mu\nu}^{(o)} + h_{\mu\nu}$ onde $h_{\mu\nu} = h_{\nu\mu}$ é uma pequena perturbação de $g_{\mu\nu}^{(o)}$. Assim, levando em conta a aproximação no lado esquerdo das Eqs. (1.1), pode-se mostrar que[2,3]

$$R_{\mu\nu} - (1/2)g_{\mu\nu} R \approx -(1/2) \partial_\lambda \partial^\lambda \psi_{\mu\nu} \qquad (2.1),$$

onde

$$\psi_{\mu\nu} = h_{\mu\nu} - (1/2)\delta_{\mu\nu} h \qquad (2.2),$$

$\delta_{\mu\nu}$ é o delta de Kronecker e h o traço de $h_{\mu\nu}$, isto é, $h = h_\alpha{}^\alpha$. Assim, usando (2.1) as (1.1) ficariam escritas como

$$\partial_\lambda \partial^\lambda \psi_{\mu\nu} = -(16\pi G/c^4) T_{\mu\nu}{}^{(m)} \qquad (2.3),$$

de onde nota−se que com a linearização perdeu−se o efeito do campo gravitacional sobre si mesmo que era levado em conta nas equações não − lineares (1.1). Devido a essa omissão a (2.3) apresenta vários defeitos: [2-6]
(a) de acordo com (2.3) a matéria age sobre o campo gravitacional (muda os campos), mas não há uma ação mútua dos campos sobre a matéria; isto é, o campo gravitacional pode adquirir energia−momento da matéria, mas a energia−momento da matéria permaneceria conservada (ref.2, §100), ou seja, obedeceria a condição $\partial^\nu T_{\mu\nu}{}^{(m)} = 0$, o que é uma inconsistência.
(b) A energia gravitacional não age como fonte de gravitação, em contradição com o princípio de equivalência.

Uma maneira de sanar tais problemas, pelo menos no limite de campos fracos,[2-6] consiste em obter um $T_{\mu\nu}$ mais geral da seguinte forma:

$$T_{\mu\nu} = T_{\mu\nu}{}^{(m)} + t_{\mu\nu} \qquad (2.4),$$

onde $t_{\mu\nu}$ deve levar em conta a energia−momento do campo gravitacional. No caso do eletromagnetismo o tensor $T_{\mu\nu}$, que leva em conta as cargas (q) e o campo eletromagnético (c) é dado por $T_{\mu\nu} = T_{\mu\nu}{}^{(q)} + T_{\mu\nu}{}^{(c)}$ pode ser visto explicitamente em muitos livros,[2,8,9] $T_{\mu\nu}{}^{(c)}$ é o *tensor de Maxwell*.

Assim, levando em conta $t_{\mu\nu}$, que é uma função de segunda ordem dos campos $h_{\mu\nu}$, como mostraremos explicitamente mais adiante, a (2.3) ficaria

$$\partial_\lambda \partial^\lambda \psi_{\mu\nu} = -(16\pi G/c^4)( T_{\mu\nu}{}^{(m)} + t_{\mu\nu}) = - (16\pi G/c^4) \tau_{\mu\nu} \qquad (2.5),$$

onde $\tau_{\mu\nu} = T_{\mu\nu}{}^{(m)} + t_{\mu\nu}$ obedece à condição $\partial^\nu \tau_{\mu\nu} = \partial^\nu (T_{\mu\nu}{}^{(m)} + t_{\mu\nu}) = 0$, ou seja, $\tau_{\mu\nu}$ é uma grandeza localmente conservada. Como $t_{\mu\nu}$ não obedece a uma lei de conservação covariante, ou seja, $t^\mu{}_{\nu;\mu} \neq 0$, ele é um *pseudo− tensor* de energia−momento (ref.2, §100). A (2.5) mostra que não só a massa inercial levada em conta em $T_{\mu\nu}{}^{(m)}$ gera o campo gravitacional $\psi_{\mu\nu}$ como também a energia e o momento contidos em $t_{\mu\nu}$.

Consideremos um corpo massivo concentrado num volume V finito em repouso num referencial S e que o efeito gravitacional das massas desse corpo seja observado num ponto P a uma distância r muito distante da origem O (vide Fig. 1).

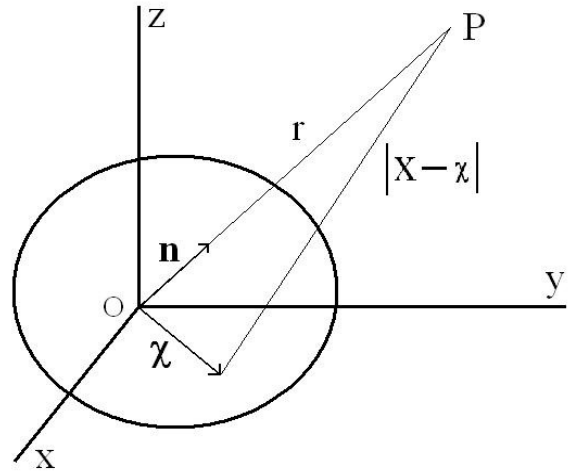

**Figura 1**

Assumindo que o campo gerado pelo corpo é muito fraco, de tal modo que temos $g_{\mu\nu} = g_{\mu\nu}^{(o)} + h_{\mu\nu}$, usando a Eq.(2.5), teremos

$$\partial_\lambda \partial^\lambda \psi_{\mu\nu}(t,\mathbf{x}) = \Box \psi_{\mu\nu}(t,\mathbf{x}) = -(16\pi G/c^4)\,\tau_{\mu\nu}(t,\mathbf{x}) \qquad (2.6),$$

onde o símbolo $\Box = \partial_\lambda \partial^\lambda = \partial^2/\partial \mathbf{x}^2 - (1/c^2)\partial^2/\partial t^2$ é o operador de d´Alembert ou "dalembertiano". A solução para tempos retardados de (2.6) é dada por [2–6,8,9]

$$\psi_{\mu\nu}(t,\mathbf{x}) = (4G/c^4) \int \tau_{\mu\nu}(t - |\mathbf{x} - \mathit{x}|/c, \mathbf{x})\, d^3\mathit{x} / |\mathbf{x} - \mathit{x}| \qquad (2.7).$$

A (2.6) e (2.7) mostram que massas e energia–momento geram campos gravitacionais $\psi_{\mu\nu}(t,\mathbf{x})$ que se propagam como ondas com a velocidade da luz. Notemos que os campos gravitacionais que se propagam são *tensoriais* ao passo que os campos eletromagnéticos são *vetoriais*.[8,9]

Uma expressão explícita e geral de $t_{\mu\nu}$, que foi obtida após um cálculo muito longo e laborioso, é mostrada na ref.2 (§100). Não nos preocuparemos em mostrá–la pois, usaremos, mais adiante, somente o caso particular de $t_{\mu\nu}$ para ondas planas.

Com uma transformação de coordenadas $x'^\mu = x^\mu + \xi^\mu$ onde $\xi^\mu$ são quantidades (vetoriais) pequenas, $g_{\mu\nu}$ fica dado por (ref.2, §94 e §101)

$$g'_{\mu\nu} = g_{\mu\nu} + \delta g_{\mu\nu} = g_{\mu\nu} - \partial^\nu \xi_\mu - \partial^\mu \xi_\nu$$

Portanto,

$$g'_{\mu\nu} = g^{(o)}_{\mu\nu} + h'_{\mu\nu} = g_{\mu\nu} - \partial^\nu \xi_\mu - \partial^\mu \xi_\nu = g_{\mu\nu}{}^{(o)} + h_{\mu\nu} - \partial^\nu \xi_\mu - \partial^\mu \xi_\nu$$

de onde obtemos

$$h'_{\mu\nu} = h_{\mu\nu} - \partial^\nu \xi_\mu - \partial^\mu \xi_\nu. \qquad (2.8).$$

Assim, como $\psi_{\mu\nu} = h_{\mu\nu} - (1/2)\delta_{\mu\nu} h$, conforme (2.2), e usando (2.8) é fácil verificarmos que

$$\partial^\mu \psi_{\mu\nu} = 0 \quad (\textit{divergência nula}) \qquad (2.9),$$

desde que os vetores $\xi_\mu$ obedeçam à equação $\Box \xi_\mu = 0$. Desse modo, sempre é possível encontrar um sistema de coordenadas $x'^\mu$ de tal modo que tenhamos satisfeita a (2.9). Consequentemente, a equação de campos (2.3) na aproximação de campos fracos fica invariante pela transformação (2.8). Isto implica que uma transformação de coordenadas $x'^\mu = x^\mu + \xi^\mu$ leva a uma transformação dos campos dada por (2.8) que mantém invariante as equações de campo definida por (2.3). Assim, a (2.9), que é conhecida como *condição de Hilbert*, expressa uma *invariância de gauge*.[2-6,7,10]

## 3. Ondas Gravitacionais Planas no Vácuo.

Antes de calcularmos o campo $\psi_{\mu\nu}$ usando a (2.7), vamos estudar como ele se propaga no vácuo. Assim, assumindo que o campo esteja se propagando no vácuo e desprezando $t_{\mu\nu}$ que contém termos de segunda ordem em $h_{\mu\nu}$, a (2.6) fica

$$\partial_\lambda \partial^\lambda \psi^{\mu\nu} = \Box \psi^{\mu\nu} = 0 \qquad (3.1).$$

Nessas condições (Apêndice III), a (3.1) se simplifica dando a *Equação de Onda*

$$\Box h_{ik} = \partial_\lambda \partial^\lambda h_{ik} = (\partial^2/\partial \mathbf{x}^2 - (1/c^2)\partial^2/\partial t^2) h_{ik} = 0 \qquad (3.2).$$

Consideremos uma onda gravitacional plana. Em tal onda o campo $h_{ik}$ varia somente em uma dada direção do espaço; escolhendo a direção de propagação ao longo do eixo $z > 0$ temos $h_{ik} = h_{ik}(t - z/c)$. Desse modo, a equação de onda (3.2) fica

$$(\partial^2/\partial z^2 - (1/c^2)\partial^2/\partial t^2) h_{ik} = 0 \qquad (3.3),$$

que tem como solução familiar

$$h_{ik}(x) = A_{ik} \cos(k_\alpha x^\alpha) \qquad (3.4),$$

onde $k_\alpha$ é o *vetor de onda* e as amplitudes $A_{ik}$ formam um tensor simétrico constante. Sendo $\omega$ a freqüência da onda, $k_4 = k_o = \omega$ e $k_2 = k_1 = 0$, $k_3 = k_z = |\mathbf{k}| = k = \omega/c$, ou seja, $k_\mu = (\omega,0,0,\omega/c)$. Levando em conta que $h_{ik}(x)$ deve obedecer (3.3), é fácil vermos que as seguintes condições devem ser obedecidas

$$A_{\beta\alpha} k^\alpha = 0 \qquad \text{e} \qquad k_\alpha k^\alpha = 0 \qquad (3.5).$$

Além dessas, com outras condições adicionais mostradas no Apêndice III, o tensor $A_{ik}$ fica dado por

$$A_{ik} = \begin{pmatrix} 0 & 0 & 0 & 0 \\ 0 & A_{11} & A_{12} & 0 \\ 0 & A_{12} & -A_{11} & 0 \\ 0 & 0 & 0 & 0 \end{pmatrix} \qquad (3.6),$$

ou seja, ele tem *traço nulo* ($A_i^i = 0$) e somente *componentes transversais* à direção z de propagação: $A_{xx} = -A_{yy}$ e $A_{xy} = A_{yx}$.

Os campos transversais $h_{xx}, h_{yy}$ e $h_{xy}$ serão representados de forma abreviada usando somente matrizes 2 x 2 chamadas de *matrizes de polarização* $(\varepsilon_+)_{ik}$ e $(\varepsilon_x)_{ik}$:

$$(\varepsilon_+)_{ik} = \begin{pmatrix} 1 & 0 \\ 0 & -1 \end{pmatrix} \qquad \text{e} \qquad (\varepsilon_x)_{ik} = \begin{pmatrix} 0 & 1 \\ 1 & 0 \end{pmatrix} \qquad (3.7),$$

Assim, a solução geral da (3.2) pode ser escrita como uma combinação linear dos campos $h_{ik}$, com *polarizações* (+) e (x), respectivamente,

$$h_{ik}^{(+)} = h_+ (\varepsilon_+)_{ik} \cos(\omega t - kz) \qquad (3.8a)$$

$$h_{ik}^{(x)} = h_x (\varepsilon_x)_{ik} \cos(\omega t - kz + \varphi) \qquad (3.8b),$$

onde $h_+ = A_{11}$, $h_x = A_{12}$ e $\varphi$ é uma fase arbitrária. A polarização tensorial das ondas gravitacionais gera um efeito muito mais complicado do que a polarização linear das ondas eletromagnéticas.

Notemos que as ondas gravitacionais $h_{ik}^{(+)}(z,t)$ e $h_{ik}^{(x)}(z,t)$ foram obtidas no limite de campos fracos e assumindo o *gauge transverso de traço nulo*[2-7].

(3.1) *Forças de Cisalhamento*.

Os campos $h_{ik}$ deformam o espaço–tempo dando origem a forças que agem sobre a matéria. Conforme mostramos no Apêndice III, os campos gravitacionais $h_{ik}^{(+)}$ e $h_{ik}^{(x)}$ criam campos de forças diferentes $\mathbf{F}^{(+)}(\mathbf{x},t)$ e $\mathbf{F}^{(x)}(\mathbf{x},t)$, respectivamente.

No caso da polarização (+) sobre uma partícula de massa m localizada no instante t = 0 no ponto (muito próximo da origem O) de coordenadas $(x_o,0,0)$ ou $(-x_o,0,0)$ atua uma força $\mathbf{F}^{(+)}$ ao longo do eixo x dada por (Apêndice III)

$$F_x^{(+)} = m(h_+/2)\, \omega^2 \cos(\omega t)\, x_o = g(t)\, x_o \qquad (3.10),$$

onde $g(t) = m(h_+/2)\, \omega^2 \cos(\omega t)$.

Analogamente, sobre uma partícula localizada em t = 0 (muito próxima da origem) no ponto com coordenadas $(0,y_o,0)$ ou $(0,-y_o,0)$, atua uma força $\mathbf{F}^{(+)}$ ao longo do eixo y é dada por

$$F_y^{(+)} = -m(h_+/2)\, \omega^2 \cos(\omega t)\, y_o = -g(t)\, y_o \qquad (3.11).$$

As (3.10) e (3.11) mostram que o campo de forças de cisalhamento devido à polarização (+), $\mathbf{F}^{(+)} = F_x^{(+)}\mathbf{i} + F_y^{(+)}\mathbf{j}$ tem uma divergência nula, ou seja, $\partial F_x^{(+)}/\partial x_o + \partial F_y^{(+)}/\partial y_o = 0$ que implica que as linhas de campos sejam hipérboles, conforme é mostrado na Fig.2.

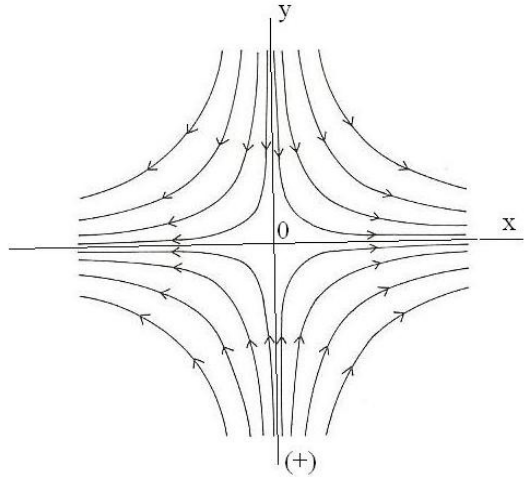

**Figura 2.** Linhas de campo de cisalhamento do tipo (+). As linhas de campo $\mathbf{F}^{(+)}$ invertem a direção em cada meio período.

No caso da polarização (x), com $\varphi = 0$, as forças sobre as partículas com as mesmas condições iniciais consideradas acima são formalmente idênticas às dadas por (3.10) e (3.11), substituindo $h_+$ por $h_x$, porém agem ao longo de eixos rodados de 45º em relação ao do referencial (x,y). Vide Figura 3.

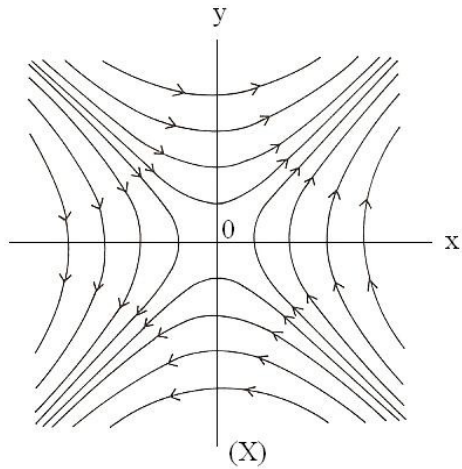

**Figura 3**. Linhas de campo de cisalhamento do tipo (x). As linhas de campo $\mathbf{F}^{(x)}$ invertem a direção em cada meio período.

Os campos $\mathbf{F}^{(+)}$ e $\mathbf{F}^{(x)}$ mostrados nas Figs.(2) e (3), são ortogonais, ou seja, $\mathbf{F}^{(+)} \cdot \mathbf{F}^{(x)} = 0$. As forças $\mathbf{F}^{(+)}(\mathbf{x},t)$ e $\mathbf{F}^{(x)}(\mathbf{x},t)$ produzem um cisalhamento dos corpos sobre os quais incidem as ondas gravitacionais. Nas Figuras 4 e 5 vemos como um colar inicialmente circular de partículas "livres" se deforma com tempo devido aos cisalhamentos (+) e (x), respectivamente.

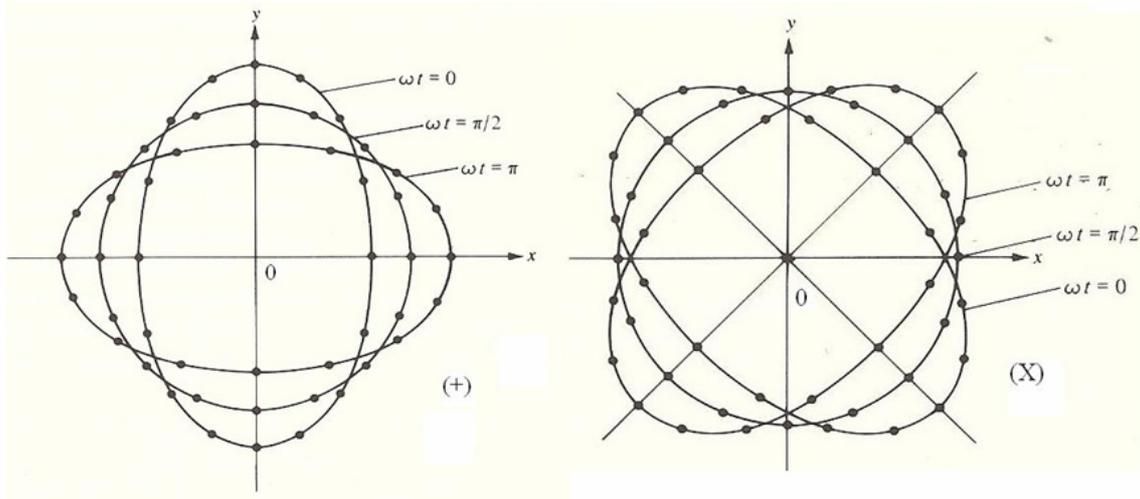

**Figuras 4 e 5**. Deformação de colares de partículas devido a ondas gravitacionais com polarizações (+) e (x), respectivamente.

Quando há simultaneamente ondas do tipo (+) e (x) o tensor $h_{ik}$ é dado pela superposição dos campos (+) e (x):

$h_{ik} = h_+ \, (\varepsilon_+)_{ik} \cos(\omega t - kz) \pm h_x \, (\varepsilon_x)_{ik} \cos(\omega t - kz + \varphi)$   (3.12),

que dá origem a ondas elipticamente polarizadas levógiras ou destrógiras, como ocorre no eletromagnetismo.[2–6,8,9] A onda *circularmente polarizada* ocorre quando $h_+ = h_x = h$ e $\varphi = 0$, vide, por exemplo, ref.6 (Cap.10.2). Usando a expressão geral para o *tensor momento angular*[1] $M_{ik}$, pode-se mostrar que as ondas planas circularmente polarizadas carregam momento angular que está orientado ao longo do eixo z, no sentido positivo ou negativo do mesmo. As primeiras têm *helicidade positiva* e as segundas *helicidade negativa*.

(3.2) *Fluxo de Energia*.

No limite de campos fracos e usando o *gauge transverso de traço nulo* o pseudo-tensor energia-momento $t_{\mu\nu}$ para ondas planas que se movem ao longo do eixo $x_3 = z$ é dado por (ref.2, §101):

$$t^{o3} = -(c^3/32\pi G)[\,(\partial_z h_{11})(\partial_t h_{11}) + (\partial_z h_{22})(\partial_t h_{22}) + 2(\partial_z h_{12})(\partial_t h_{12})] \qquad (3.13).$$

Notemos que $t^{o3}$ é um termo de segunda ordem em $h_{ik}$. Como os campos $h_{ik}$ são funções somente de $t - z/c$, pode-se mostrar que (ref.2, §101)

$$t^{o3} = (c^2/16\pi G)\{(\dot{h}_{12})^2 + (1/4)[\dot{h}_{11} - \dot{h}_{22}]^2\} \qquad (3.14),$$

onde o ponto sobre a letra significa derivada no tempo.

As ondas gravitacionais transportando energia criam em torno delas um campo de gravitação. Este campo é de um infinitésimo de ordem superior ao do campo da onda que gera uma energia, que como vemos em (3.14), já é muitíssimo pequeno em segunda ordem de $h_{ik}$.

Como $h_{11} = -h_{22}$ de (3.14) obtemos

$$t^{o3} = (c^3/16\pi G)[(\dot{h}_{12})^2 + (\dot{h}_{11})^2] = (c^3/16\pi G)[(\dot{h}_x)^2 + (\dot{h}_+)^2] \qquad (3.15),$$

onde ambas as polarizações com $h_+ = h_{11} = -h_{22}$ e $h_x = h_{12} = h_{21}$ estão sendo consideradas para o fluxo de energia. Assim, o fluxo de energia $\Phi_z = ct^{o3}$ ao longo da direção z, fica escrita na forma

$$\Phi_z = ct^{o3} = (c^3/32\pi G)[(\dot{h}_{12})^2 + (\dot{h}_{11})^2 + (\dot{h}_{21})^2 + (\dot{h}_{22})^2] =$$

$$= (c^3/32\pi G)(\dot{h}_{ij}\dot{h}_{ij}) \qquad (3.16).$$

Um modo mais simples e instrutivo para obter a (3.16) pode ser visto na ref.6 (Apêndice E).

Fazendo uma média no tempo, usando (3.10), obtemos o *fluxo médio de energia* $<\Phi_z> = c <t^{o3}>$, medido em energia/(área.s), ao longo do eixo z, devido às ondas com polarização (+) e (x):

$$<\Phi_z> = (c^3/32\pi G)(h_+^2 + h_x^2)\omega^2 \qquad (3.17).$$

Pode-se mostrar (ref.5, Cap.4.1) que a razão entre quantidade de momento angular transportada por uma onda circularmente polarizada e a energia transportada por ela é igual a $\pm(2/\omega)$. O sinal + é para a helicidade positiva e o sinal − para a helicidade negativa. Assumindo que os grávitons transportem a energia, cada um com energia $\hbar\omega$ e momento angular $S\hbar$ teríamos a razão (momento angular/energia) dada por $S\hbar/\hbar\omega$, de onde pode-se concluir que S = 2, ou seja, os que grávitons teriam spin 2. Os spins S = 2 estariam, então, orientados ao longo da direção de propagação da onda com $S_z = \pm 2$.

## 4. Ondas Gravitacionais Quadrupolares.

Suponhamos que os efeitos gravitacionais sejam observados a uma distância r da origem O (vide Fig.1) tal que $|\mathbf{x}| = r \gg |\mathbf{x}|$, como ocorre na zona de radiação do eletromagnetismo. Assim, da (2.7) resulta

$$\psi_{\mu\nu}(t,\mathbf{x}) = (4G/c^4 r) \int T_{\mu\nu}(t - r/c, \mathbf{x}) d^3\mathbf{x} \qquad (4.1).$$

Para simplificar a notação e levando em conta que temos uma função do tempo retardado $t - r/c$, poremos simplesmente $T_{\mu\nu}(t - r/c, \mathbf{x}) = \tau_{\mu\nu}$. Assim, a (4.1) fica,

$$\psi_{\mu\nu}(t,\mathbf{x}) = (4G/c^4 r) \int \tau_{\mu\nu} d^3\mathbf{x} \qquad (4.2).$$

Como $\partial^\nu T_{\mu\nu}^{(m)} = 0 = \partial^\nu \tau_{\mu\nu}$ as componentes espaciais e temporais ficam dadas por:

$$\partial \tau_{\alpha o}/\partial x^o = \partial \tau_{\alpha\gamma}/\partial x^\gamma \quad \text{e} \quad \partial \tau_{o\gamma}/\partial x^\gamma = \partial \tau_{oo}/\partial x^o \qquad (4.3).$$

Multiplicando a primeira equação por $x^\beta$ e integrando no volume em todo o espaço teremos, para o lado esquerdo e direito,

$$\int (\partial \tau_{\alpha o}/\partial x^o) x^\beta dV = (\partial/\partial x^o) \int \tau_{\alpha o} x^\beta dV \quad \text{e}$$

$$\int (\partial \tau_{\alpha\gamma}/\partial x^\gamma) x^\beta dV = \int \partial(\tau_{\alpha\gamma} x^\beta)/\partial x^\gamma dV - \int \tau_{\alpha\beta} dV = 0 - \int \tau_{\alpha\beta} dV,$$

usando o teorema de Gauss e levando em conta que $\tau_{\alpha\gamma}$ se anula no infinito. Assim,

$$\int \tau_{\alpha\gamma}\, dV = - (\partial/\partial x^o)\int \tau_{\alpha o}\, x^\beta\, dV \ .$$

Como $\tau_{\alpha\beta}$ é simétrico, fazendo a semi−soma da equação acima obtemos,

$$\int \tau_{\alpha\gamma}\, dV = -(1/2)(\partial/\partial x^o)\int [\tau_{\alpha o}\, x^\beta + \tau_{\beta o}\, x^\alpha]\, dV \qquad (4.4).$$

Agora, multiplicando a segunda equação de (4.3) por $x^\alpha x^\beta$ e integrando em todo o espaço e procedendo de modo semelhante para obter a (4.4) teremos

$$(\partial/\partial x^o)\int \tau_{oo}\, x^\alpha\, x^\beta\, dV = -\int [\tau_{\alpha o}\, x^\beta + \tau_{\beta o}\, x^\alpha]\, dV \qquad (4.5).$$

Comparando (4.4) e (4.5) vemos que

$$\int \tau_{\alpha\gamma}\, dV = (1/2)\, (\partial^2/\partial x_o^2)\int \tau_{oo}\, x^\alpha\, x^\beta\, dV \qquad (4.6).$$

Desprezando $t_{\mu\nu}$ que contém somente termos de segunda ordem em $h_{\mu\nu}$, levando em conta que o efeito preponderante de $T_{\mu\nu}^{(m)}$, conforme (1.4), é dado pela densidade de energia própria $\varepsilon$ das partículas em V e que as suas velocidades v sejam muito menores do que c poremos $T_{oo} \approx \varepsilon\, u_o u_o \approx \varepsilon = \rho_o c^2$, lembrando que para v/c << 1 temos $u_o = u_4 = \gamma \approx 1$ (Apêndice II). Nessas condições, como $x_o = x_4 = ct$, usando a Eq.(4.6) vemos que o tensor $\psi_{\mu\nu}(t,\mathbf{x})$ definido por (4.2) fica

$$\psi_{\alpha\beta}(t,\mathbf{x}) = (2G/c^4 r)(\partial^2/\partial t^2)\int \varepsilon\, x^\alpha x^\beta\, dV = (2G/c^2 r)(\partial^2/\partial t^2)\int \rho_o\, x^\alpha x^\beta\, dV \qquad (4.7),$$

lembrando que o lado direito da (4.7) é calculado para um tempo retardado t − r/c. Usando o *tensor momento de quadrupolo* $Q_{\alpha\beta}$ das massas em S definido por

$$Q_{\alpha\beta} = \int \rho_o\, (3x^\alpha x^\beta - \delta_{\alpha\beta}\, r^2)\, dV \qquad (4.8),$$

a (4.7) torna−se

$$\psi_{\alpha\beta}(t,\mathbf{x}) = (2G/c^2 r)(\partial^2 Q_{\alpha\beta}/\partial t^2) = (2G/c^2 r)\ddot{Q}_{\alpha\beta} \qquad (4.9),$$

onde os pontos sobre a letra significam derivadas em relação a t.

A grandes distâncias do corpo emissor que está em S, podemos considerar, numa pequena região da zona de radiação, a onda como sendo plana. Assim, podemos calcular o fluxo de energia emitido pelo sistema ao longo, por exemplo, do eixo z = $x^3$ usando a (3.14) levando em conta que $h_{12} = \Psi_{12}$ e $h_{11} - h_{22} = \Psi_{11} - \Psi_{22}$. Assim, teremos

$$h_{12} = (2G/c^2 r)\ddot{Q}_{12} \quad e \quad h_{11} - h_{22} = (2G/c^2 r)[\ddot{Q}_{11} - \ddot{Q}_{22}] \qquad (4.10).$$

Desse modo usando (3.14) e (4.10), o fluxo de energia $\Phi_z = ct^{o3}$ ao longo do eixo z = $x^3$ é dado por

$$\Phi_z = ct^{o3} = (G/36\pi c^5 r^2) \{(\ddot{Q}_{12})^2 + [\ddot{Q}_{11} - \ddot{Q}_{22}]^2/4\} \qquad (4.11).$$

Conhecendo o fluxo ao longo do eixo z = $x^3$, é fácil determinar ao fluxo ao longo do versor genérico **n** = $n_1$ **i** + $n_2$ **j** + $n_3$ **k** (vide Fig.1) que está ao longo do vetor de posição **r** = r **n**; (**i,j,k**) são os versores ao longo dos eixos x, y e z, respectivamente, no sistema de coordenadas onde o corpo emissor está em repouso; $n_1$ = sinθ cosφ, $n_2$ = sinθ sinφ e $n_3$ = cosθ, onde os ângulos θ e φ são os ângulos polares de **n**. Assim, intensidade de radiação dI = dE/dt que passa por uma área d**A** = dA **n** = $r^2$dΩ **n**, localizada no ponto r, é dada por (ref.2, pág.426)

$$dI = (G/36\pi c^5) \{(1/2)\ddot{Q}_{\alpha\beta}^2 - \ddot{Q}_{\alpha\beta}\ddot{Q}_{\alpha\gamma} n_\beta n_\gamma + (1/4)(\ddot{Q}_{\alpha\beta} n_\alpha n_\beta)^2\} d\Omega \qquad (4.12),$$

onde dΩ é o elemento de ângulo sólido.

A energia irradiada em todas as direções, ou seja, a *energia perdida* pelo sistema emissor por unidade de tempo (− dE/dt) pode ser obtida (ref.2,§104) fazendo uma média do fluxo dado por (4.12) em todas as direções e multiplicando essa média por 4π :

$$- dE/dt = (G/45c^5) \ddot{Q}_{\alpha\beta}^2 \qquad (4.13).$$

**Apêndice I. Tempo próprio τ.**

Consideremos dois relógios, R e R´, fixos em dois referenciais inerciais S e S´ que movem com velocidade relativa v. Em S temos $ds^2 = c^2 dt^2 - dx^2 - dy^2 - dz^2$ e em S´ temos $ds'^2 = c^2 dt'^2 - dx'^2 - dy'^2 - dz'^2$. No intervalo de tempo dt´ medido pelo relógio R´ que está fixo em S´, portanto, com dx´= dy´= dz´ = 0, temos ds´ = cdt´= $cdx_4$´. Observado de S o R´se deslocou de uma distância d**r** = (dx,dy,dz) = $\{dx_i\}_{i=1,2,3}$ num intervalo de tempo dt. Como $ds^2 = ds'^2$ obtemos

$$dt' = ds/c = [c^2 dt^2 - dx^2 - dy^2 - dz^2]^{1/2}/c = dt [1 - (dx^2 + dy^2 + dz^2)/c^2 dt^2]^{1/2},$$

ou seja,

$$dt' = ds/c = dt\,[1 - (v/c)^2]^{1/2} \qquad (I.1),$$

onde $v = [(dx^2+dy^2+dz^2)/dt^2]1/2$ é a velocidade do relógio móvel R´ em relação ao referencial S. O tempo t´, medido no referencial S´ onde R´ está em repouso, é denominado de *tempo próprio* τ. Desse modo, segundo a (I.1), o intervalo de *tempo próprio* dτ é dado por

$$d\tau = ds/c = dt\,[1 - (v/c)^2]^{1/2} \qquad (I.2).$$

Segundo (I.2), o tempo τ medido (tempo próprio) por um observador que anda junto com um relógio é menor do que o tempo observado de referencial que se move com velocidade v relativamente ao relógio.

### Apêndice II. Quadrivetor velocidade $u_\alpha$.

No caso geral o quadrivetor *velocidade* $u_\alpha$ de uma partícula é definido por $u_\alpha = dx_\alpha/ds$. Num espaço de Minkowski onde temos

$$ds = cdt\,[1 - (v/c)^2]^{1/2} \qquad (II.1),$$

v é a velocidade ordinária da partícula em 3−dim dada por **v** = (dx/dt, dy/dt, dz/dt) = $\{dx_i/dt\}_{i=1,2,3}$ = $\{v_i\}_{i=1,2,3}$. Definindo o fator $\gamma = 1/[1 - (v/c)^2]^{1/2}$ e usando (II.1) o quadrivetor velocidade $u_\alpha = (u_4, u_i)$ fica dado por,

$$u_4 = u_o = dx_4/ds = cdt/ds = \gamma \quad \text{e}$$

$$u_i = dx_i/ds = (dx_i/dt)\gamma/c = (v_i/c)\gamma \quad (i=1,2,3), \qquad (II.2).$$

Assim, como $u_\alpha = (u_4, u_i) = (\gamma, v_i\gamma/c)$ vemos que $u_\alpha u^\alpha = (u_\alpha)^2 = g_{\alpha\beta} u_\alpha u^\beta$ é dado por

$$u_\alpha^2 = 1 \qquad (II.3),$$

mostrando que $u_\alpha$ é um *quadrivetor unidade*.

### Apêndice III. Ondas Gravitacionais Planas no Vácuo.
**(III.1)** *Equação de Onda.*

Para pequenos $h_{ik}$, as componentes $\Gamma^i{}_{kl}$, dadas por (1.3) que se exprimem por derivadas dos $g_{ik}$, são também pequenas. Desprezando potências superiores à primeira ordem em $h_{ik}$, o *tensor de curvatura* $R_{iklm}$ ou de *Riemann−Christoffel* [1−6] fica dado por (ref.2, §101)

$$R_{iklm} \approx \{(\partial^2 h_{im}/\partial x^k \partial x^l) + (\partial^2 h_{kl}/\partial x^i \partial x^m) - (\partial^2 h_{km}/\partial x^i \partial x^l) - (\partial^2 h_{il}/\partial x^k \partial x^m)\}\,.$$

Assim, o *tensor de Ricci* [1–6] $R_{ik} = g^{lm} R_{iklm}$ com a mesma precisão temos $R_{ik} \approx g_{(o)}^{lm} R_{iklm}$, ou seja, (ref.2, §101)

$$R_{ik} \approx \{- g_{(o)}^{lm} (\partial^2 h_{ik}/\partial x^l \partial x^m) + (\partial^2 h_i^l/\partial x^k \partial x^l)$$
$$+ (\partial^2 h_k^l/\partial x^i \partial x^l) - (\partial^2 h/\partial x^i \partial x^k)\} \qquad (III.1.1).$$

Partindo das equações fundamentais de Einstein (1.1) para componentes mistas,

$$R_i^k - (1/2)\delta_i^k R = \kappa T_i^{k\,(m)}$$

e efetuando uma contração sobre os índices i e k, ou seja, fazendo $\delta_i^k R_i^k - (1/2) \delta_i^k \delta_i^k R = \kappa \delta_i^k T_i^{k\,(m)}$, obtemos $R = -\kappa T^{(m)}$, onde $T^{(m)} = T_i^{i(m)}$. Isso mostra que as equações de campo podem ser também escritas na forma

$$R_{ik} = \kappa \{T_i^{k\,(m)} - (1/2)g_{ik}T^{(m)}\} \qquad (III.1.2).$$

Como no vácuo $T_{ik}^{(m)} = 0$ s equações de campo (III.1.2) se reduzem a

$$R_{ik} = 0 \qquad (III.1.3).$$

Levando em conta que $\psi_{ik} = h_{ik} - (1/2)\delta_{ik} h$ obedece à *condição de gauge* $\partial^i \psi_{ik} = 0$, verificamos que $R_{ik}$ dada pela (III.1.1) torna-se

$$R_{ik} \approx -(1/2)\Box h_{ik} \qquad (III.1.4).$$

Consequentemente, verificamos, usando (III.1.3) e (III.1.4) que as equações do campo no vácuo assumem a forma

$$\Box h_{ik} = 0 \qquad (III.1.5).$$

(III.2) *Tensor de Polarização $A_{ik}$.*

No caso de uma onda plana se propagando no sentido positivo do eixo $z = x^3$, a equação de onda $\Box h_{ik} = 0$ para $h_{ik} = h_{ik}(t-z/c)$ fica escrita

$$(\partial^2/\partial z^2 - (1/c^2)\partial^2/\partial t^2) h_{ik} = 0 \qquad (III.2.1),$$

que tem como solução familiar

$$h_{ik}(x) = A_{ik} \cos(k_\alpha x^\alpha) \qquad (III.2.2),$$

onde $k_\alpha$ é o *vetor de onda* as amplitudes $A_{ik}$ formam um tensor simétrico constante. Sendo $\omega$ a freqüência da onda, $k_4 = k_o = \omega$ e $k_2 = k_1 = 0$, $k_3 = k_z = |\mathbf{k}| = k = \omega/c$, ou seja, $k_\mu = (\omega,0,0,\omega/c)$.

Usando as *equações de gauge* ( *divergência nula*) (2.9) e levando em conta que os campos $\psi_{ik} = \psi_{ik}(\zeta)$ são só funções de t e z, pois $\zeta = \zeta(t - z/c)$, obtemos $\partial^1\psi_{i1} = \partial^2\psi_{i2} = 0$ e $\partial^o\psi_{io} = \partial^3\psi_{i1}$. Como $\psi_{ik} = \psi_{ik}(\zeta)$ e $x_o = ct$ resulta $\partial^t\psi_{io} = \partial^t\psi_{i3}$ que, integrando no tempo dão, colocando a constante de integração igual a zero pois só função do tempo vai contribuir,

$$\psi_{io} = \psi_{i1} \ (i = 0,1,2,3) \qquad (III.2.3),$$

ou seja, $\psi_{1o} = \psi_{11}$, $\psi_{2o} = \psi_{21}$, $\psi_{3o} = \psi_{31}$ e $\psi_{oo} = \psi_{o1}$ (ref.2, §101).

Com uma transformação *adequada* de coordenadas $x'_i = x_i + \xi_i(t-z/c)$ podemos anular 4 elementos de $\psi_{ik}$ que escolheremos como sendo $\psi_{o1}$, $\psi_{o2}$, $\psi_{o3}$ e $\psi_{22} + \psi_{33}$. Assim, levando em conta também (III.2.3) e que $\psi_{ik} = \psi_{ki}$, teremos

$$\psi_{oi} = \psi_{io} \ (i = 0,1,2,3) \quad e \quad \psi_{11} = \psi_{oo} = 0 \ e \ \psi_{22} + \psi_{33} = 0 \qquad (III.2.4).$$

Da última relação (III.2.4) resulta que o traço de $\psi_{ik}$ é nulo, ou seja, $\psi_i^i = 0$. Como $\psi_{ik} = h_{ik} - (1/2)\delta_{ik} h$ resulta $h = h_i^i = 0$ e, portanto,

$$\psi_{ik} = h_{ik} \qquad (III.2.5).$$

Resumindo, usando os resultados mostrados acima, temos as seguintes equações para $h_{ik}$

$$h_\alpha^\alpha = h = 0 \ (\textit{traço nulo}) \qquad (III.2.6a)$$

$$h_{\alpha o} = 0 \qquad (\alpha = 0,1,2,3) \qquad (III.2.6b)$$

e

$$\partial^\alpha h_{\alpha\beta} = 0 \quad (\textit{divergente nulo}) \qquad (III.2.6c)$$

Levando em conta que $h_{ik}(x)$, dada por (III.2.2), deve obedecer à equação de onda (III.2.1)) e às Eqs.(III.3.6), é fácil vermos que as seguintes condições devem ser obedecidas

$$A_{\beta\alpha} k^\alpha = 0 \qquad e \qquad k_\alpha k^\alpha = 0 \qquad (III.2.7).$$

Lembrando que $k_2 = k_1 = 0$, $k_4 = k_o = \omega$ e $k_3 = k = \omega/c$ obtemos de (III.2.7) $A_{\beta o} \omega + A_{\beta 3} \omega/c = 0$ e impondo ainda a condição (III.2.6b) dada por $h_{\beta o} = 0 = A_{\beta o}$ ($\beta = 0,1,2,3$) obtemos $A_{\beta 3} = 0$ para todos os $\beta$. Como $h_{ik}$ é simétrico decorre $A_{\beta 3} = A_{3\beta} = A_{\beta o} = A_{o\beta} = 0$ e também que $A_{12} = A_{21}$. Com a condição de *traço nulo* teremos $A_{11} + A_{22} = 0$ de onde tiramos $A_{22} = - A_{11}$.

Assim, o tensor $A_{ik}$ fica dado por

$$A_{ik} = \begin{pmatrix} 0 & 0 & 0 & 0 \\ 0 & A_{11} & A_{12} & 0 \\ 0 & A_{12} & -A_{11} & 0 \\ 0 & 0 & 0 & 0 \end{pmatrix} \quad \text{(III.2.8)},$$

ou seja, ele tem *traço nulo* ($A_i^i = 0$) e somente *componentes transversais* à direção de propagação z da onda tais que $A_{xx} = - A_{yy}$ e $A_{xy} = A_{yx}$.

Resultados semelhantes aos mostrados acima para $A_{ik}$ podem ser obtidos usando a invariância de gauge,[7] definida por (2.8).

(III.3) *Forças de cisalhamento*.

Vamos primeiramente calcular as distâncias $\Delta\ell$ em 3−dim entre duas partículas com massa m localizadas muito próximas da origem O (vide Figs.4 e 5) ao longo dos eixos x e y, levando em conta a deformação do espaço−tempo devido a ondas com polarização (+).

Consideremos primeiramente duas partículas localizadas ao longo do eixo x (y=z=0), uma no ponto $x = - x_o$ e a outra em $x = x_o$. Notemos que x são as coordenadas dos pontos. A *distância* $\Delta\ell_x$ entre elas (para campos fracos), usando (3.8a), é dada por

$$\Delta\ell_x^2 = g_{11}\Delta x^2 = (1 - h_{11})(2x_o)^2 = (1 - h_+ \cos \omega t)(2x_o)^2 \quad \text{(III.3.1)},$$

ou seja,

$$\Delta\ell_x \approx [1 - (h_+/2)\cos(\omega t)](2x_o) \quad \text{(III.3.2)}.$$

Analogamente, se as partículas estão ao longo do eixo y (x=z=0) com coordenadas $y = - y_o$ e $y = y_o$, a distância $\Delta\ell_y$ entre elas será dada por

$$\Delta\ell_y \approx (g_{22})^{1/2}\Delta y = [1 + (h_+/2)\cos(\omega t)](2y_o) \quad \text{(III.3.3)}.$$

Usando (III.3.2) e (III.3.3) as acelerações de cada partícula em direção à origem ao longo de x e de y são dadas, respectivamente, por $d^2(\Delta\ell_x)/dt^2 = (\omega^2 h_+/2)\cos(\omega t)x_o$ e $d^2(\Delta\ell_y)/dt^2 = -(\omega^2 h_+/2)\cos(\omega t)y_o$.

Assim, as forças $F_x^{(+)}$ e $F_y^{(+)}$ sobre uma partícula ao longo dos eixos x e y são dadas, respectivamente, por

$$F_x^{(+)} = [(m\omega^2 h_+/2)\cos(\omega t)] x_o = g(t) x_o \quad (III.3.4a)$$

e

$$F_y^{(+)} = -[(m\omega^2 h_+/2)\cos(\omega t)] y_o = -g(t) y_o \quad (III.3.4b).$$

As relações (III.3.4a,b) mostram que o campo de forças de cisalhamento devido à polarização (+), $\mathbf{F}_{(+)} = F_x^{(+)} \mathbf{i} + F_y^{(+)} \mathbf{j}$ tem uma divergência nula, ou seja, $\partial F_x^{(+)}/\partial x_o + \partial F_y^{(+)}/\partial y_o = 0$ o que implica que as linhas de campos sejam hipérboles, conforme é mostrado na Fig.2. É fácil vermos que a componente radial $F_r^{(+)}$ das forças de cisalhamento $\mathbf{F}_{(+)}$ é dada por

$$F_r^{(+)} = [(m\omega^2 h_+/2)\cos(\omega t)] r_o \cos(2\theta),$$

onde $r_o = (x_o^2 + y_o^2)^{1/2}$ e $\theta$ é o ângulo azimutal.

O campo de forças $\mathbf{F}_{(x)}$, visto na Fig.3, devido à polarização (x) é obtido de modo análogo ao adotado para a polarização (+). A única diferença é que as linhas de forças hiperbólicas no caso (x) estão rodadas de um ângulo de $\theta = 45°$ em relação ao eixo x.